\documentclass[preprintnumbers,amsmath,amssymb,aps,prl,floatfix,lengthcheck,superscriptaddress]{revtex4}
\usepackage{graphicx}
\usepackage{braket}
\usepackage{upgreek}
\usepackage{subscript}
\usepackage{color}

\newcommand{\rubidium}{\textsuperscript{87}Rb}
\newcommand{\us}{$\upmu$s}
\newcommand{\um}{$\upmu$m}
\newcommand{\uK}{$\upmu$K}
\newcommand{\density}[2]{#1$\times$10\textsuperscript{#2}\unit{}{cm\textsuperscript{-3}}}
\newcommand{\unit}[2]{\mbox{#1\,#2}} 
\newcommand{\trapfreq}[2]{$\omega$\textsubscript{#1}\,=\,2$\pi$$\times$\unit{#2}{Hz}}

\begin{document}

\title{Photo-association of trilobite Rydberg molecules via resonant spin-orbit coupling}

\author{K. S.\ Kleinbach}
\author{F. Meinert}
\author{F. Engel}
\author{W. J. Kwon}
\author{R. L\"{o}w}
\author{T. Pfau}
\affiliation{5. Physikalisches Institut and Center for Integrated Quantum Science and Technology, Universit\"{a}t Stuttgart, Pfaffenwaldring 57, 70569 Stuttgart, Germany}
\author{G. Raithel}
\affiliation{Department of Physics, University of Michigan, Ann Arbor, Michigan 48109, USA}
\date{\today}

\begin{abstract}
We report on a novel method for photo-association of strongly polar trilobite Rydberg molecules.
This exotic ultralong-range dimer, consisting of a ground-state atom bound to the Rydberg electron via electron-neutral scattering, inherits its polar character from the admixture of high angular momentum electronic orbitals.
The absence of low-$L$ character hinders standard photo-association techniques.
Here, we show that for suitable principal quantum numbers resonant coupling of the orbital motion with the nuclear spin of the perturber, mediated by electron-neutral scattering, hybridizes the trilobite molecular potential with the more conventional ${\rm{S}}$-type molecular state. This provides a general path to associate trilobite molecules with large electric dipole moments, as demonstrated via high-resolution spectroscopy. We find a dipole moment of 135(45)\,D for the trilobite state. Our results are compared to theoretical predictions based on a Fermi-model.

\end{abstract}

\maketitle

Controlling dipolar molecules at ultralow temperatures is of central interest for a plethora of developments and applications in ultracold chemistry \cite{Ospelkaus2010}, precision spectroscopy, quantum information processing \cite{DeMille2002}, or quantum many-body physics \cite{Carr2009}.
In the case of ultracold Rydberg gases, the observation of a novel type of molecule formed by a neutral ground-state atom trapped within the giant Rydberg electron wavefunction \cite{Bendkowsky2009,Greene2000,Fabrikant2002} has recently sparked intense interest in their few- and many-body properties \cite{Gaj2014,Balewski2013,Schmidt2016}. 
Astonishingly, these ultralong-range Rydberg molecules can possess large permanent electric dipole moments, which arise from the coupling of many high-angular momentum Rydberg states \cite{Greene2000, Li2011, Shaffer2015,Niederpruem2016b} by electron-neutral scattering. 
While dimers correlated with ${\rm{S}}$- \cite{Bendkowsky2009,Tallant2012,DeSalvo2015}, ${\rm{P}}$- \cite{Sassmannshausen2015}, and ${\rm{D}}$- \cite{Anderson2014b,Krupp2014} Rydberg states have been efficiently photo-associated from atomic ground-state ensembles via single- or two-photon excitation pathways, they typically inherit relatively weak high-$L$ admixture, owing to the large non-integer parts of their quantum defects, and consequently exhibit comparatively small dipole moments, e.g. $\sim 1$ Debye for Rb \cite{Li2011}. 
In contrast, the dipole moment of exotic trilobite states associated with the quasi-degenerate hydrogenic Rydberg manifolds may reach kilo-Debye dipole moments \cite{Greene2000}. 
Yet, these states are generally not accessible to standard photo-association schemes due to selection rules, unless favorable conditions are met, such as the presence of Rydberg $\rm{S}$-states that are almost degenerate with a high-$L$ manifold \cite{Shaffer2015}. 
For low principal quantum numbers, trilobite-like states have also been reached starting from pre-associated weakly-bound dimers \cite{Stwalley2013}.

In this Letter, we demonstrate a novel method for photo-association of trilobite Rydberg molecules using $^{87}$Rb as an example.
The method is generally applicable for a multitude of atomic species.
Our scheme relies on the recently explored coupling of electron-neutral singlet and triplet scattering channels mediated by the ground-state hyperfine interaction.
While in  low-$L$ Rydberg dimers this causes molecular states of mixed spin character \cite{Sassmannshausen2015,Anderson2014,Boettcher2016,Niederpruem2016}, here we show that the angular momentum couplings lead to hybridization of the trilobite state with S-state molecules around principal quantum number $n\approx50$, where the energy gap between high-$L$ and optically accessible low-$L$ states matches the hyperfine splitting.
The hybrid character of these trilobite molecules provides large permanent dipole moments and offers access via two-photon laser-association.

We consider diatomic ultralong-range Rydberg molecules formed by the interaction of the Rydberg electron with a neutral perturber atom in its electronic ground state. 
The interaction is expressed in terms of a Fermi contact potential that accounts for $s$- and $p$-wave electron-neutral scattering \cite{Fermi1934,Omont1977}.
 In order to capture the coupling between trilobite and S-state molecules, we must take into account the relevant spin degrees of freedom and their mutual couplings. 
The Hamiltonian describing the full spin system and the Rydberg electron at distance $R$ between the Rydberg ionic core and the ground-state perturber reads \cite{Greene2000,Anderson2014,AU}
\begin{align}
\hat{H}(\mathbf{r},R) =& \hat{H}_0 + \sum\limits_{i=S,T} 2 \pi a_{s}^{i}(k) \delta^3(\mathbf{r}-R \hat{\mathbf{z}}) \hat{\mathcal{P}}_i \nonumber \\
& + \sum\limits_{i=S,T} 6 \pi a_{p}^{i}(k) \delta^3(\mathbf{r}-R \hat{\mathbf{z}}) \overleftarrow{\nabla} \cdot \overrightarrow{\nabla} \hat{\mathcal{P}}_i \nonumber \\
&+ A_{\rm{hfs}} \hat{\mathbf{S}}_2 \cdot \hat{\mathbf{I}}_2 \, . 
\label{eq1}
\end{align}
Here, $\hat{H}_0$ is the Hamiltonian of the unperturbed Rydberg electron including fine structure coupling between orbital and spin angular momentum $\hat{\mathbf{L}}_1$ and $\hat{\mathbf{S}}_1$ via published quantum defects \cite{Mack2011,Li2003,Han2006}. 
The second and third term describe the $s$-wave and $p$-wave interaction, parametrized by an energy-dependent scattering length $a_{s,p}^{S,T}(k)$, where $k$ is the position-dependent electron momentum in the scattering process, and $S$ ($T$) denotes singlet (triplet) configuration of the coupled Rydberg and ground-state atom electron spin $\hat{\mathbf{S}}_1$ and $\hat{\mathbf{S}}_2$. 
Accordingly, $\hat{\mathcal{P}}_T = \hat{\mathbf{S}}_1 \cdot \hat{\mathbf{S}}_2 + 3/4$ and $\hat{\mathcal{P}}_S = 1 - \hat{\mathcal{P}}_T$ are the projection operators onto singlet and triplet states.
The internuclear axis is pointing along the $\mathbf{z}$-direction and $\mathbf{r}$  denotes the spatial coordinate of the Rydberg electron.
The hyperfine coupling between $\hat{\mathbf{S}}_2$ and the nuclear spin $\hat{\mathbf{I}}_2$ of the perturber atom has a strength of $A_{\rm{hfs}}$ \cite{Anderson2014}.

\begin{figure}[!ht]
\centering
	\includegraphics[width=\columnwidth]{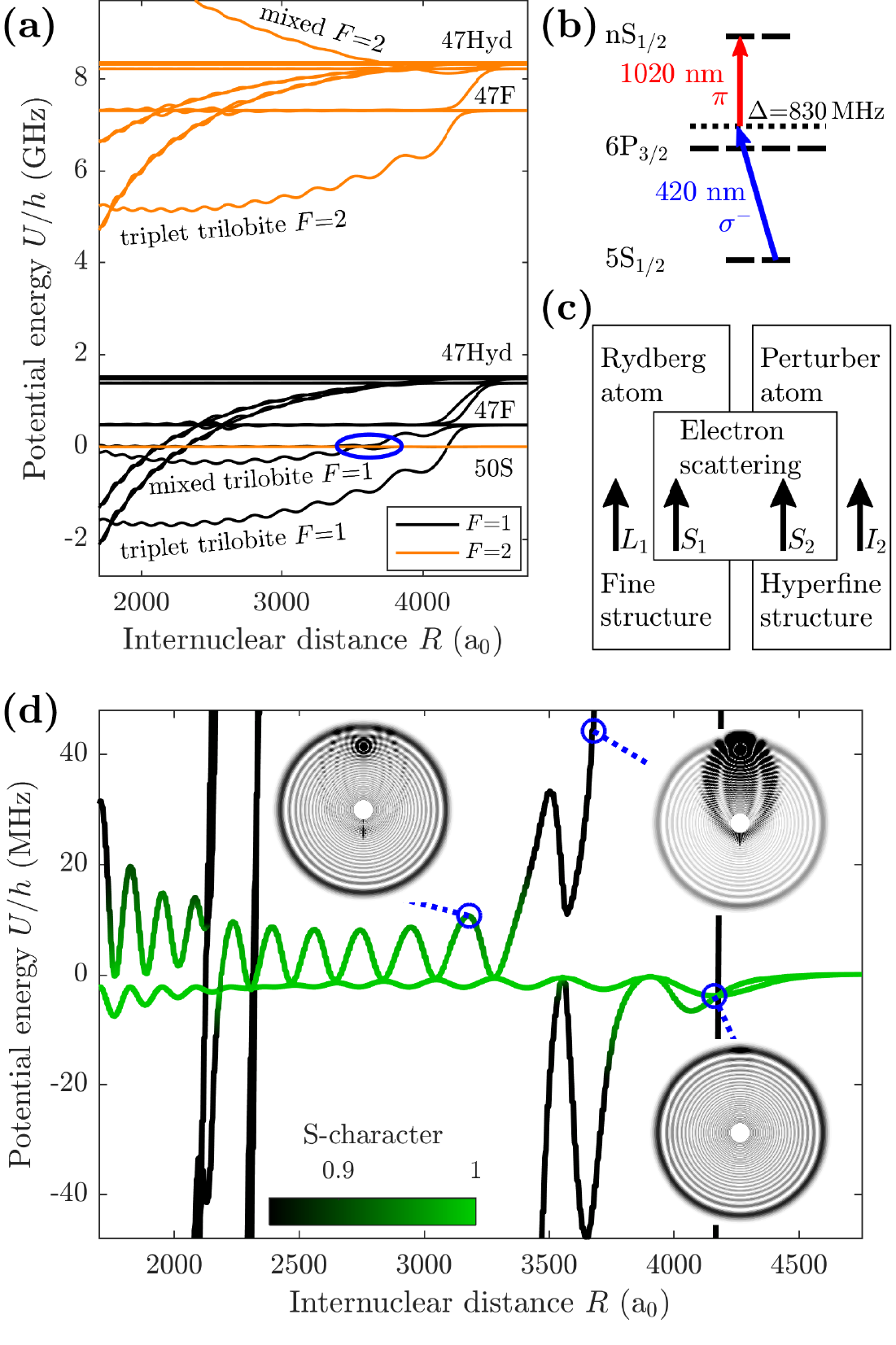}
	\caption{(a) Long-range molecular potential energy $U$ for $^{87}$Rb as a function of the internuclear distance $R$ between the Rydberg core and the perturber atom in the vicinity of the $50\rm{S}$ Rydberg state. Zero energy corresponds to the asymptotic pair state $50{\rm{S}}, F=2$. The energy gap to the next hydrogenic manifold $47{\rm{Hyd}}, F=2$ is comparable to the hyperfine splitting between $F=2$ and $F=1$ of $\sim 6.8$ GHz.
	(b) Schematic of the two-photon laser excitation employed in the experiment for coupling to $n\rm{S}$ Rydberg states.
	(c) Illustration of angular momentum couplings for Rydberg molecule potentials. Electron-neutral scattering couples the electron spins of the Rydberg and perturber atom. In combination with fine and hyperfine structure, the pure singlet and triplet scattering channels and the hyperfine states of the perturber get mixed.
	(d) Close-up showing the resulting hybridization of the $47{\rm{Hyd}}, F=1$ singlet trilobite potential with the $50{\rm{S}}, F=2$ molecular potential. The green shading indicates the {\rm{S}}-character of the PECs. Insets depict the Rydberg electron density for three selected values of $R$ marked with blue circles.
	}
	\label{fig:Potential}
\end{figure}

Diagonalizing the Hamiltonian Eq.~\ref{eq1} for each internuclear distance $R$ yields a set of  Born-Oppenheimer molecular potential energy curves (PECs) \cite{Bendkowsky2009,Greene2000}, which we show in Fig.~\ref{fig:Potential}(a) for our diatomic $^{87}$Rb system in the vicinity of the $50\rm{S}_{1/2}$ Rydberg state \cite{SM}.
We start the discussion with the PECs near the $n=47$ hydrogenic manifold and with the perturber atom prepared in the upper hyperfine level of total angular momentum $F=2$ (orange curves). 
With decreasing $R$ the $s$-wave scattering term strongly mixes the nearly degenerate hydrogenic Rydberg states of different $L$ and causes a single PEC (labeled triplet trilobite $F=2$) to detach from the manifold.
This PEC supports deeply bound molecular states exhibiting Rydberg electron densities which are strongly localized in the vicinity of the perturber atom with a characteristic shape resembling a trilobite fossil \cite{Greene2000}.
As a consequence, such trilobite Rydberg molecules possess extraordinary large permanent electric dipole moments. 
Yet, their high-$L$ character hampers direct photo-association via single- or two-photon processes, unless favorable atomic-species specific properties exist.
For instance a practically integer $\rm{S}$-state quantum defect  in Cs causes sizeable admixture of $\rm{S}$-state character in trilobite states \cite{Shaffer2015}.

\begin{figure*}
\centering
	\includegraphics[width=\textwidth]{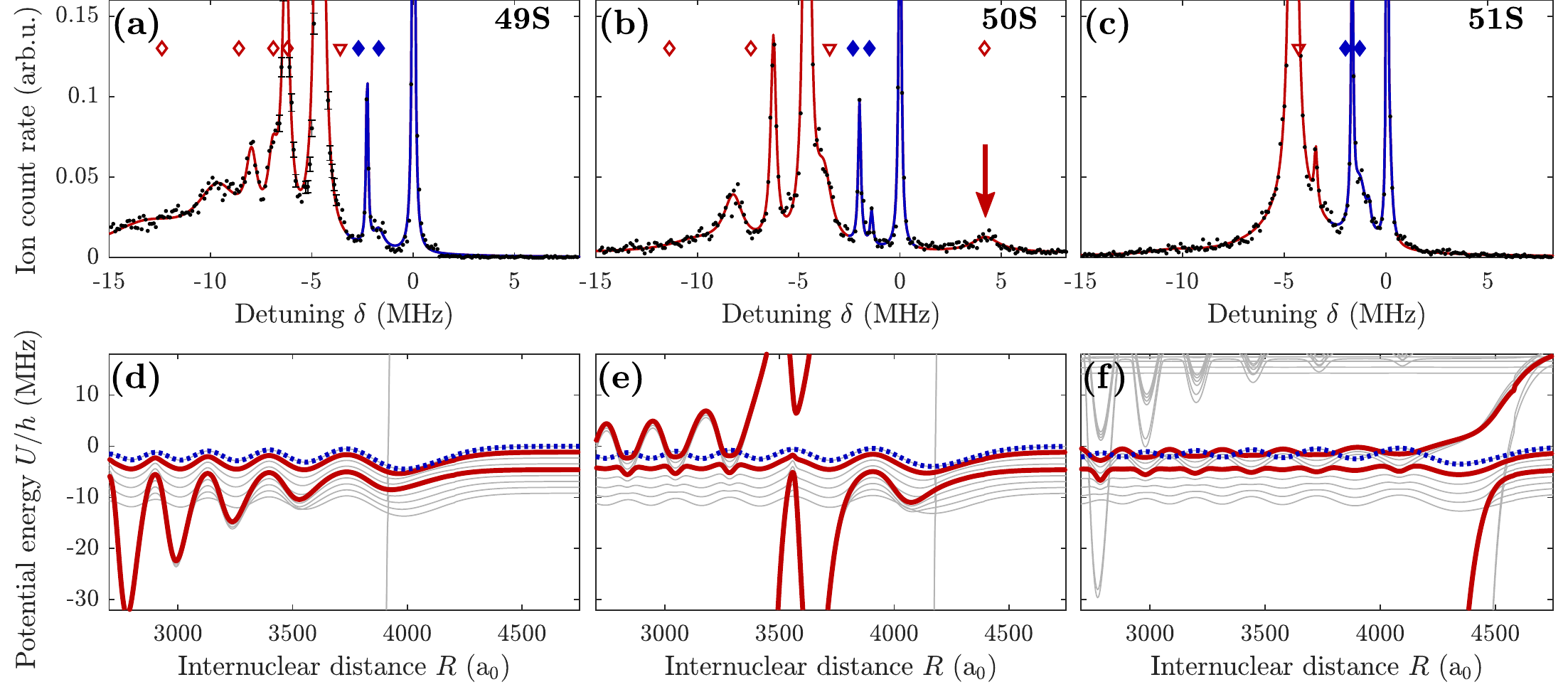}
	\caption{
	(a)-(c) Rydberg molecule spectra showing the mean ion count rate as a function of laser detuning $\delta$ in the vicinity of the atomic Rydberg states $49\rm{S}$ (a), $50\rm{S}$ (b), and $51\rm{S}$ (c) at a magnetic offset field of \unit{1.64}{G}. 
	The zero of the frequency axis ($\delta=0$) is referenced to the $\ket{n{\rm{S}}_{1/2},\uparrow}$ atomic Rydberg level. 
	Representative error bars in (a) indicate the standard deviation. Solid lines are fits to the data based on a sum of multiple Lorentzians. 
	Blue filled (red open) symbols indicate calculated binding energies of molecular states associated with PECs for which $m_K=5/2$ ($m_K=3/2$). 
	(d)-(f) Long-range molecular potentials in the vicinity of the Rydberg states $49\rm{S}$ (d), $50\rm{S}$ (e), and $51\rm{S}$ (f) for the experimental magnetic field of \unit{1.64}{G}. 
	Blue dotted (red solid) curves show PECs with $m_K=5/2$ ($m_K=3/2$). 
	Zero energy corresponds to the asymptotes of the states $n{\rm{S}}, F=2$ with $m_K=5/2$.
	}
	\label{fig:3Spectra_Pot}
\end{figure*}
In contrast, our method to associate trilobite Rydberg molecules is based solely on the angular momentum couplings involved in the Hamiltonian Eq.~\ref{eq1} (Fig.~\ref{fig:Potential}(c)) and thus applicable to a wide class of elements.
Due to simultaneous hyperfine and electron-neutral interaction, the full Hilbert space can in general no longer be separated into subspaces of well defined hyperfine ($F=1,2$) and singlet/triplet scattering states.
More specifically, the hyperfine coupling in the perturber atom mixes the pure singlet and triplet scattering channels \cite{Anderson2014,Boettcher2016}. 
Vice versa, the electron-neutral interaction may generally cause mixing of the unperturbed $F=1$ and $F=2$ hyperfine levels \cite{Niederpruem2016}.
The amount of mixing depends on the relative strengths of the interaction terms and the detunings of nearby states, which we exploit in the following to control the hybridization of trilobite and S-type molecular potentials by tuning $n$. The strongest coupling appears for the upper trilobite PEC detaching from the $n=47, F=1$ manifold (labeled mixed trilobite $F=1$), which crosses the PEC attached to the $50{\rm{S}}, F=2$ asymptote around $R\sim 3500 \, a_0$, where $a_0$ is the Bohr radius (blue circle in Fig.~\ref{fig:Potential}(a)). 
A close-up of the avoided crossing induced by the combined effects of hyperfine and electron-neutral interaction is shown in Fig.~\ref{fig:Potential}(d) together with representative plots of the orbital Rydberg electron density, which show how the hybridization varies as a function of $R$. 
For an internuclear distance well beyond the avoided level crossing the molecular potential correlated with the $50{\rm{S}}, F=2$ asymptote exhibits a clean $\rm{S}$-type electron density. 
Yet, with decreasing $R$ the Rydberg electron acquires the characteristic trilobite shape and eventually forms a hybrid of $\rm{S}$- and mixed high-$L$ character.
The sizeable S-character allows us to photo-associate molecular bound states associated with the mixed trilobite PEC.

Starting from an ultracold sample (\unit{1.2}{\uK}) of $^{87}\rm{Rb}$ held in a magnetic QUIC trap and prepared in the \mbox{$|F=2,m_F=2\rangle$} hyperfine state, we couple to $n\rm{S}_{1/2}$ Rydberg states by a two-photon transition (see Fig.~\ref{fig:Potential}(b)) \cite{SM}.
For the chosen laser polarizations and the detuning from the intermediate $6\rm{P}_{3/2}$ level, both Zeeman sublevels $\ket{n{\rm{S}}_{1/2},\uparrow}$ and $\ket{n{\rm{S}}_{1/2},\downarrow}$ are addressed, where \mbox{$\uparrow(\downarrow)$} denotes the Rydberg electron spin projection \mbox{$m_{J_1}=+1/2(-1/2)$}.
We take Rydberg molecule spectra by sequential laser excitation, field ionization, and ion detection as detailed in \cite{SM}. Results of such measurements showing the mean ion yield as a function of laser detuning $\delta$, referenced to the atomic Rydberg level $\ket{n{\rm{S}}_{1/2},\uparrow}$, are shown in Figs.~\ref{fig:3Spectra_Pot}(a)-(c) in the vicinity of the states $49\rm{S}$, $50\rm{S}$ and $51\rm{S}$.
The transition to the Rydberg state with flipped electron spin $\ket{n{\rm{S}}_{1/2},\downarrow}$ is Zeeman-shifted to $\delta=\unit{-4.6}{MHz}$ due to the magnetic offset field $\mathbf{B}$ present in the QUIC trap.
In addition to the two atomic Rydberg lines, we observe further narrow resonances red detuned with respect to the atomic transitions, which are attributed to photo-association of ultralong-range Rydberg molecules.
Remarkably, the spectrum in the vicinity of the $50\rm{S}_{1/2}$ level (Fig.~\ref{fig:3Spectra_Pot}(b)) reveals a bound molecular state at blue detuning $\delta=\unit{+4.2}{MHz}$ (red arrow). Similar blue-detuned resonances were observed in \cite{Tallant2012}; these are, however, caused by a hybridization mechanism of near-degenerate low- and high-$L$ orbital Rydberg states that does not involve a spin-orbit interaction.

In order to interpret and assign the observed spectral lines, our theoretical model Eq.~\ref{eq1} is extended to account for the presence of $\mathbf{B}$  \cite{SM}.
We have verified that the angle between the internuclear axis and $\mathbf{B}$ has only minor influence and thus restrict the discussion to a parallel alignment \cite{SM}.
In that case, the full Hamiltonian conserves the total spin projection quantum number $m_K=m_{J_1}+m_{S_2}+m_{I_2}$.
The obtained PECs in the vicinity of the experimentally investigated Rydberg states are shown in Figs.~\ref{fig:3Spectra_Pot} (d)-(f).
In our experiment we only address PECs with $m_K=\{3/2,5/2\}$, as $m_{J_1}=\pm 1/2$ and the initially spin polarized sample fixes $m_{S_2}=1/2$ and $m_{I_2}=3/2$.
The PEC with $m_K=5/2$ (blue dotted lines) is of pure triplet character and couples to neither of the $F=1$ trilobite potentials.
In contrast, PECs with $m_K=3/2$ (red solid lines) are strongly influenced by the coupling to the $F=1$ trilobite state of mixed singlet and triplet character.
The strength of this coupling varies considerably for different $n$.
For the $49{\rm{S}}$ state (Fig.~\ref{fig:3Spectra_Pot}(d)), the trilobite potential stays blue detuned for all values of $R$ and does not cross at all. Yet, despite its comparatively large energy gap of more than \unit{50}{MHz} to the $49{\rm{S}}$ state, the coupling pushes the $m_K=3/2$ PEC significantly down in energy.
In contrast, for $50{\rm{S}}$ (Fig.~\ref{fig:3Spectra_Pot}(e)) the trilobite state crosses the ${\rm{S}}$-state at about $3500 \, a_0$ and strong hybridization with the $m_K=3/2$ PEC arises. The hybrid character even remains for smaller values of $R$, leading to a potential curve which supports bound molecular states blue detuned with respect to the atomic line, in agreement with our experimental observation.
Remarkably, such blue-detuned molecular lines do not exist for other types of ultralong-range $^{87}$Rb Rydberg molecules and support direct evidence for the strong coupling to the triblobite PEC.
Finally, for the $51{\rm{S}}$ state the crossing with the trilobite potential occurs at a larger internuclear distance of about $4400 \, a_0$ and has again less influence on the ${\rm{S}}$-state. 

Next, we calculate the molecular bound vibrational states for the PECs of interest and compare the results with the measured resonance positions.
For this, numerically obtained binding energies are indicated in Figs.~\ref{fig:3Spectra_Pot}(a)-(c).
The binding energies of the ground and first-excited dimer-state associated with the $m_K=5/2$ PEC (blue filled diamonds) are found in good agreement with the experimental data.
Furthermore, the general trend of the calculated bound states in the $m_K=3/2$ potentials (red open symbols) agrees with our measurements. 
While molecular states of the energetically higher $m_K=3/2$ PEC (triangles) appear blue detuned with respect to the $\ket{n{\rm{S}}_{1/2},\downarrow}$ atomic line, the energetically lower $m_K=3/2$ PEC supports bound states (diamonds) which are all located on the red side of the atomic resonance.
Our numerical results signify that the latter are considerably influenced by the coupling to the trilobite potential.
Note that for $51{\rm{S}}$ the absence of ion signal red detuned with respect to the $\ket{n{\rm{S}}_{1/2},\downarrow}$ atomic line perfectly follows the numerical predictions. 
The unusual blue-detuned ($\delta>0$) molecular state found in the $50{\rm{S}}$ spectrum is also predicted by our theoretical model.

\begin{figure}
\centering
	\includegraphics[width=\columnwidth]{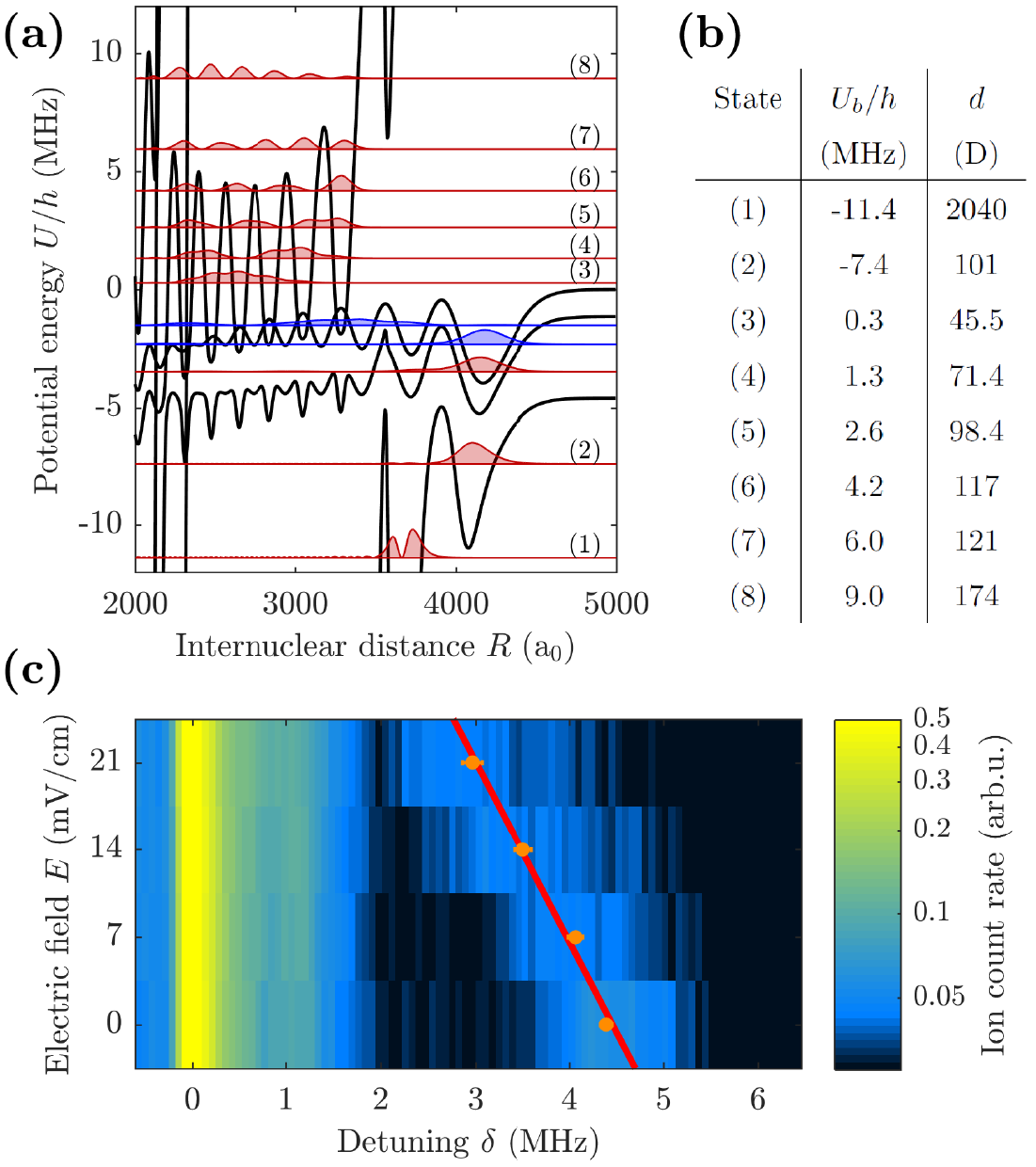}
	\caption{
	(a) Long-range molecular potentials with $m_K=\{3/2,5/2\}$ in the vicinity of the $50{\rm{S}}$ state and calculated vibrational bound states $|\Phi_v(R)|^2$, offset by their respective binding energies $U_b$. The red (blue) shaded states are bound in the PEC with $m_K=3/2$ $(m_K=5/2)$, respectively.
	Bound states with significant trilobite admixture are numbered (1)-(8). Zero energy corresponds to the asymptotes of the states $50{\rm{S}}, F=2$ with $m_K=5/2$.
	(b) Calculated binding energies $U_b$ and dipole moments $d$ for the vibrational bound states (1)-(8) in (a). (c) Rydberg Stark map showing the ion count rate as a function of the laser detuning $\delta$ in the vicinity of the $50{\rm{S}}$ state for four values of the electric field $E$. The zero of the frequency axis ($\delta=0$) is referenced to the atomic Rydberg level $\ket{n{\rm{S}}_{1/2},\uparrow}$ in the field-free case. Orange points depict the peak position of the blue detuned trilobite molecular state extracted from Lorentzian fits to the data. The red line is a linear fit to the peak positions, which yields an electric dipole moment of \unit{135(45)}{D}.
	}
	\label{fig:StarkMap}
\end{figure}

We now focus on the hybrid PEC with sizeable trilobite character, which is energetically located above the $\ket{50{\rm{S}}_{1/2},\uparrow}$ atomic Rydberg level and hosts the peculiar blue-shifted molecular state.
The relevant PECs together with their vibrational bound states $|\Phi_v(R)|^2$ are depicted in Fig.~\ref{fig:StarkMap}(a).
Vibrational states associated with the hybrid trilobite potentials are labeled \mbox{(1)-(8)}.
The corresponding calculated binding energies $U_b$ are listed in Fig.~\ref{fig:StarkMap}(b).
The measured binding energy of $\unit{4.2}{MHz}$ is found in agreement with the calculated state (6).
To provide further evidence for this assignment, we probe the electric dipole moment $d$ of the molecule and compare to theoretical predictions (\textit{cf.} Fig.~\ref{fig:StarkMap}(b)) \cite{SM}.
To this end, we record photo-association spectra while applying a homogeneous electric field $E$ pointing parallel to the QUIC trap magnetic field \cite{SM}.
The result of this measurement, shown in Fig.~\ref{fig:StarkMap}(c), reveals a pronounced linear red shift of the resonance position by several MHz together with a small broadening for applied electric fields of only a few mV/cm.
As expected, while the molecular line is strongly affected, the atomic resonance exhibits no detectable Stark shift for these comparatively small field strengths. 

The linear Stark shift of the molecular resonance attests a large permanent electric dipole moment $d=h\times d\delta/dE=$\unit{135(45)}{D}. 
This observation unambiguously indicates the trilobite character of the molecular state.
Moreover, the extracted dipole moment as well as the measured binding energy are both in good agreement with the theoretical prediction for state (6).
This identification is further supported by a reasonably long calculated lifetime $\tau=\unit{1.8}{\us}$ and the fact that the associated vibrational wavefunction has comparatively large weight at large $R\sim$ \unit{3300}{a$_0$}. 
This facilitates efficient photo-association for our typical inter-particle distances when compared to the other blue-shifted vibrational states, which are all located at smaller expectation values of $R$.

Finally, we note that conventional {\rm{S}}-type ultralong-range Rydberg molecules, which feature only a small perturbative admixture of far detuned high-$L$ states, typically show a symmetric broadening of the molecular line relative to the atomic transition in response to electric fields \cite{Li2011}.
This differs from what we observe here and hints at the need for a more elaborate theoretical description.

In conclusion, we have studied the hybridization of \mbox{${\rm{S}}$-type} and trilobite Rydberg molecular potentials induced by the combined effects of electron-neutral scattering and hyperfine interaction.
 The coupling allows us to produce exotic polar trilobite Rydberg molecules, related to high angular momentum electron orbits, via standard two-color photo-association. The strong linear Stark shift observed in the presence of an electric field provides direct evidence for their large permanent electric dipole moment. Our results exemplify how fine details of the chemical bond of ultralong-range Rydberg dimers, emerging from the full interplay of orbital and spin-degrees of freedom, allow us to access molecular states which are otherwise difficult to produce. Importantly, our scheme is readily adaptable to numerous atomic species that feature hyperfine structure. This opens routes for tailored engineering of long-range interacting few- to many-body systems based on strongly polar Rydberg molecules \cite{Eiles2017}, realizing exotic polarons \cite{Schmidt2016,Schmidt2016b}, or probing quantum chemistry on mesoscopic scales \cite{Schlagmueller2016}.

We thank S. Hofferberth and F. B\"ottcher for fruitful discussions and I. I. Fabrikant for providing scattering phase shifts. We acknowledge support from Deutsche Forschungsgemeinschaft (DFG) within the SFB/TRR21 and the project PF 381/13-1, and the NSF (Grant No. PHY-1506093).

\newpage
\clearpage

\section{Supplementary Material: Photo-association of trilobite Rydberg molecules via resonant spin-orbit coupling}

\subsection{Calculation of Born-Oppenheimer potential energy curves (PECs)}

For calculating Born-Oppenheimer potential energy curves, we evaluate the matrix elements of $\hat{H}$ for fixed values of $R$ in a basis set $\{ |n,L_1,J_1,m_{J_1} ; m_{I_2}, m_{S_2} \rangle \}$, where $n$ and $L_1$ label principal and orbital angular momentum quantum numbers of the Rydberg electron.
The total spin-orbit coupled angular momentum of the Rydberg state and its projection along $\mathbf{z}$ are denoted with $J_1$ and $m_{J_1}$, respectively. Finally, $m_{I_2}$ and $m_{S_2}$ are the projections of the nuclear and electronic spin of the ground-state perturber.
We note that the hyperfine splitting of the Rydberg levels is below the resolution of our experiment and can therefore be neglected.
For all presented PECs, the basis set for diagonalization is chosen to comprise a span of four hydrogenic manifolds \cite{Fey2015MAT,Eiles2016MAT}, explicitly ranging from $n=45$ to $n=48$ for the case of 50S.
The energy-dependent scattering lengths relate to the corresponding phase shifts $\delta_{s,p}^{S,T}(k)$ via $a_s^{S,T}(k)  = -\tan(\delta_s^{S,T}(k))/k$ and $a_p^{S,T}(k)  = -\tan(\delta_p^{S,T}(k))/k^3$ \cite{Fabrikant1986MAT}.
The electron momentum in the scattering process $k$ depends on the internuclear distance $R$ and is obtained using the semiclassical expression for the Rydberg electron's kinetic energy $k(R)^2/2 = -1/(2 n^{\star 2}) + 1/R$, where $ n^\star$ is the effective principal quantum number of the Rydberg level of interest \cite{Greene2000MAT}.

\subsection{Orbital angular momentum of the trilobite molecule}

\begin{figure}
\centering
	\includegraphics[width=\columnwidth]{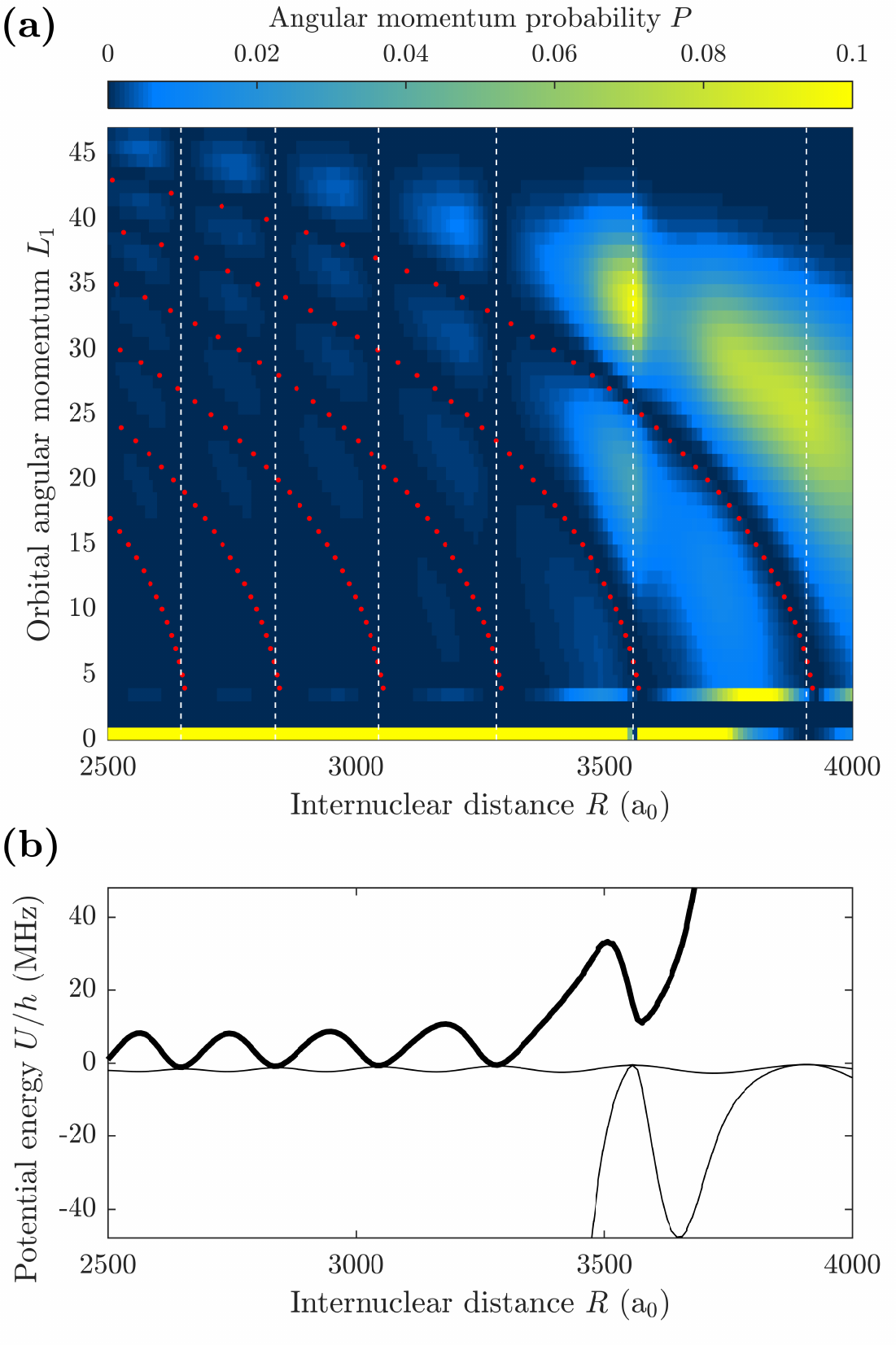}
	\caption{Orbital angular momentum character of the hybridized $^{87}$Rb trilobite molecule. (a) Angular-momentum distribution $P(L_1)$ as a function of the internuclear distance $R$ for the hybridized trilobite molecule in the vicinity of the 50S Rydberg state. The white dashed lines indicate the values of $R$ for which the 50S-wavefunction has a node. The red dotted lines show the nodes of the high-$L$ wavefunctions. (b) Molecular potential energy $U$ as a function of $R$ in the vicinity of the 50S Rydberg state. The thick line shows the potential energy curve for which $P(L_1)$ is plotted in (a) (\textit{cf.} Fig. 1 of the main text).
	}
	\label{fig:S1L_character}
\end{figure}

The diagonalization which yields the Born-Oppenheimer potential energy curves also provides the corresponding eigenstates for each value of $R$. This allows us to compute the Rydberg electron densities shown in Fig.~1 of the main article. Moreover, the angular-momentum probability distribution of the Rydberg electron, $P(L_1)$, can be extracted for all Born-Oppenheimer adiabatic states, as a function of the internuclear separation $R$. In Fig.~\ref{fig:S1L_character} we show $P(L_1)$ for the adiabatic potential that exhibits strong hybridization between the 50S state and the relevant $n=47$ trilobite state. These results illustrate the range of high-$L$ orbital angular momentum states that contribute in shaping the characteristic electron wavefunction of the trilobite. The straight, vertical node lines in Fig.~\ref{fig:S1L_character}(a) occur at fixed positions $R$ at which the hybridization vanishes for all $L_1$ because the 50S-wavefunction has a node at those $R$-values. Conversely, the curved node lines reflect values of $L_1$ and $R$ for which the hybridization vanishes because the high-$L$ wavefunction has a node. We have also verified that the major contribution to the hybrid trilobite potential comes from states with principal quantum numbers $n=50$ and $n=47$.

\subsection{Molecular potential energy curves in the presence of magnetic field}

For the numerical data shown in Fig.~2 of the main article, we have added the Zeeman term describing the interaction with the external magnetic field $\mathbf{B}$
$$
\hat{H}_B = \frac{\mathbf{B}}{2} \boldsymbol{\cdot}  (\hat{\mathbf{L}}_1+g_S\hat{\mathbf{S}}_1+g_S\hat{\mathbf{S}}_2+g_I\hat{\mathbf{I}}_2).
$$
to the Hamiltonian before diagonalizing the full system. Here, $g_S$ ($g_I$) is the electron (nuclear) spin $g$-factor.
The magnetic field splits the PECs correlated with the $n{\rm{S}}_{1/2}, F=2$ asymptote into ten lines associated with magnetic projections $m_F=\{-2,-1,...,+1,+2\}$ and $m_{J_1}=\pm 1/2$, from which two are energetically degenerate.

In our measurements, we do not resolve the weak rotational level splitting (typically $\sim$kHz) of the Rydberg molecules.
As a consequence, the location of the ground-state atom is essentially random for each experimental realization.
This generally requires to average the calculated spectra over all possible alignment angles between the internuclear axis and the magnetic-field direction in order to model the experiment.
Yet, we have found through our numerical calculations that the alignment angle between the field direction and the internuclear axis has only minor influence on all PECs relevant to this work.
In calculations that allow for an arbitrary alignment angle, we have seen that for S-type Rydberg states the PECs are independent of that angle.
This is expected, because the S-type molecules are of Hund’s case (b). In this case, the system of intrinsic spins - $S_1$, $S_2$, and $I_2$ - couples independently of the orbital $L_1=0$ state.
The spin system aligns with the magnetic field, while the orbital state is essentially isotropic and has no alignment.
For the trilobite wavefunctions the molecular binding is very strong and decouples the orbital and intrinsic spins, leaving the latter to independently align with the magnetic field, while the trilobite wavefunction is locked to the internuclear axis.
Therefore, the mixed S- and trilobite states in this work also do not depend on the alignment angle between internuclear axis and magnetic field.

\subsection{Trilobite Rydberg molecule spectroscopy}

Our experiments demonstrating photo-excitation of trilobite Rydberg molecules start from an ultracold sample of 6.6$\times$10\textsuperscript{6} \rubidium{} atoms in the $\ket{5{\rm{S}}_{1/2},F=2,m_F=2}$ state, prepared at a temperature of about \unit{1.2}{\uK}.
The cloud is held in a magnetic QUIC trap \cite{Esslinger1998MAT} with trap frequencies \trapfreq{r}{200} in the radial and \trapfreq{ax}{15} in the axial direction, resulting in a peak density of \density{2.0}{12}.
We employ a two-photon transition involving the intermediate $6{\rm{P}}_{3/2}$ level to couple the electronic ground state to $n\rm{S}_{1/2}$ Rydberg states.
For this, the sample is illuminated simultaneously with two frequency-stabilized laser beams at wavelengths $\lambda_1=\unit{420}{nm}$ and $\lambda_2=\unit{1020}{nm}$ focused to waists of \unit{2}{mm} and \unit{2.1(3)}{\um}, respectively.
The laser polarizations are set to $\sigma^-$ for the \unit{420}{nm} beam and $\pi$ for the \unit{1020}{nm} beam, relative to the QUIC-trap magnetic field.
For sufficiently large single-photon detuning $\Delta$ from the intermediate state $\ket{6{\rm{P}}_{3/2}}$, much larger than its hyperfine splitting, this couples the spin polarized electronic ground state to the $m_{J_1}=-1/2$ Rydberg level, for which the electron spin orientation has flipped during excitation (Fig.~1(b) in the main article).
Yet, in our experiment the single-photon detuning is set to $\Delta = $ \unit{830}{MHz}, which causes residual coupling to the $m_{J_1}=+1/2$ state. 
The magnetic offset field of \unit{1.64}{G} pointing along the axial direction of our QUIC trap allows us to resolve both $m_J$ levels.
We find an about four times larger excitation probability for the $m_{J_1}=-1/2$ component.
In order to avoid any residual Rydberg-Rydberg interaction, we carefully adjust the laser parameters to keep the average number of created Rydberg excitations well below unity.
After illuminating the cloud for \unit{20}{\us}, a strong electric field of \unit{360}{V/cm} is applied in order to ionize the Rydberg atoms and guide the produced ions towards a channeltron detector.
This process of excitation, ionization and subsequent detection is typically repeated 1000 times at a rate of \unit{2}{kHz} for each single atomic sample.
As a consequence, the atom number decreases by $\sim$$25\%$, accompanied by a modest increase of the cloud temperature to \unit{1.4}{\uK}.
To obtain Rydberg molecule spectra, presented in Fig.~2 of the main article, we repeat the aforementioned measurement sequence for different frequency detunings $\delta$ of the \unit{1020}{nm} laser.
Each data point is then an average over three such experimental runs. 

\subsection{Measurement of the electric dipole moment}

For the measurement of the electric field dependence of the trilobite Rydberg molecule (Fig.~3(c) in the main article), the experimental procedure as detailed in the previous section is adapted to avoid any influence of possible residual field drifts. 
Specifically, we now take a full scan of $\delta$ employing two successive frequency ramps in opposite directions and repeat this sequence 20 times in a single atomic sample.
This is done consecutively for four increasing values of $E$.
The procedure is repeated 125 times, and the results are averaged.
Compared to the previously applied sequence, the number of laser excitations performed in a single sample is augmented to 4000.
This results in an increased final temperature of  \unit{2.5}{\uK} and decreased atom number of 4.5$\times$10\textsuperscript{6} at the end of the 4000 excitations.

\subsection{Calculation of the electric dipole moment}
The electric dipole moments $d$ for the molecular states listed in Fig.~3(b) of the main article are given by
$$
d = e \int |\Phi_v(R)|^2 \bra{\Psi_{el}(R)}\hat{z}\ket{\Psi_{el}(R)} dR \, .
$$
Here, $e$ denotes the elementary charge and $\Psi_{el}(R)$ is the adiabatic electron/spin state at location $R$ on the PEC of interest, which is obtained from our full diagonalization of $\hat{H}$.
The $R$-dependent expectation values of the electronic dipole operator along the $\mathbf{z}$-direction are weigthed with the square modulus of the vibrational wavefunction of interest $\Phi_v(R)$.

\clearpage


\begin{references}
\bibitem{AU} We give the Hamiltonian in atomic units.
\end{references}
\end{document}